\documentclass[aps,prc,twocolumn,superscriptaddress,floatfix,amsmath,amssymb,nofootinbib,longbibliography]{revtex4-2}
\usepackage[utf8]{inputenc}
\usepackage[T1]{fontenc}

\usepackage{xcolor}
\usepackage{graphicx}
\usepackage{microtype}
\usepackage{braket}
\usepackage{isotope}
\usepackage{cancel}

\usepackage{orcidlink}

\usepackage{hyperref}
\hypersetup{
    colorlinks = true,
    citecolor  = blue,
    linkcolor  = blue,
    urlcolor  = blue
}

\newcommand{\ai}{\textit{ab initio}}
\newcommand{\ie}{i.e.}
\newcommand{\eg}{e.g.}

\newcommand{\la}{\langle}
\newcommand{\ra}{\rangle}
\newcommand{\psitgt}{|\Psi(\vec{c}_\circ) \ra}

\newcommand{\elem}[2]{$^{#2}${#1}}
\newcommand{\MeV}{\text{MeV}}
\newcommand{\fm}[1]{\ensuremath{\text{fm}^{#1}}}
\newcommand{\NEC}{\ensuremath{N_\text{EC}}}
\newcommand{\nxlo}[1]{N$^{#1}$LO}

\newcommand{\nuhamil}{\texttt{NuHamil}}
\newcommand{\imsrg}{\texttt{imsrg++}}

\begin{document}

\title{Hartree-Fock emulators for nuclei: Application to charge radii of $^{48,52}$Ca}

\author{M.~Companys Franzke \orcidlink{0009-0001-8623-5235}}
\email{margarida.companys@tu-darmstadt.de}
\affiliation{Technische Universität Darmstadt, Department of Physics, 64289 Darmstadt, Germany}
\affiliation{ExtreMe Matter Institute EMMI, GSI Helmholtzzentrum für Schwerionenforschung GmbH, 64291 Darmstadt, Germany}
\affiliation{Max-Planck-Institut für Kernphysik, Saupfercheckweg 1, 69117 Heidelberg, Germany}

\author{A.~Tichai \orcidlink{0000-0002-0618-0685}}
\email{alexander.tichai@tu-darmstadt.de}
\affiliation{Technische Universität Darmstadt, Department of Physics, 64289 Darmstadt, Germany}
\affiliation{ExtreMe Matter Institute EMMI, GSI Helmholtzzentrum für Schwerionenforschung GmbH, 64291 Darmstadt, Germany}
\affiliation{Max-Planck-Institut für Kernphysik, Saupfercheckweg 1, 69117 Heidelberg, Germany}

\author{K.~Hebeler \orcidlink{0000-0003-0640-1801}}
\email{kai.hebeler@tu-darmstadt.de}
\affiliation{Technische Universität Darmstadt, Department of Physics, 64289 Darmstadt, Germany}
\affiliation{ExtreMe Matter Institute EMMI, GSI Helmholtzzentrum für Schwerionenforschung GmbH, 64291 Darmstadt, Germany}
\affiliation{Max-Planck-Institut für Kernphysik, Saupfercheckweg 1, 69117 Heidelberg, Germany}

\author{A.~Schwenk \orcidlink{0000-0001-8027-4076}}
\email{achim.schwenk@tu-darmstadt.de}
\affiliation{Technische Universität Darmstadt, Department of Physics, 64289 Darmstadt, Germany}
\affiliation{ExtreMe Matter Institute EMMI, GSI Helmholtzzentrum für Schwerionenforschung GmbH, 64291 Darmstadt, Germany}
\affiliation{Max-Planck-Institut für Kernphysik, Saupfercheckweg 1, 69117 Heidelberg, Germany}

\begin{abstract}
Understanding the emergence of complex structures of nuclei from chiral effective field theory (EFT) is a central challenge. The large number of low-energy couplings (LECs) in the EFT expansion and the significant cost of \ai{} many-body calculations render large-scale sensitivity studies of many-body observables computationally prohibitive, necessitating the use of emulators as low-cost surrogates. In this work, we study a Hartree-Fock emulator based on eigenvector continuation to investigate trends in nuclear charge radii of neutron-rich calcium isotopes. We systematically vary the five LECs entering the leading three-nucleon ($3N$) interactions, and demonstrate the precision of the emulator through cross validation over a wide parameter space. Our findings indicate that the large charge-radius increase from \elem{Ca}{48} to \elem{Ca}{52} is likely not explained by variations of the leading $3N$ couplings. This suggests that other effects, such as sensitivities to chiral two-nucleon interactions or neglected many-body effects, \eg{}, associated with nuclear collectivity, play an important role.
\end{abstract}

\maketitle

\section{Introduction}

{\it Ab initio} calculations of nuclei combine a systematic expansion of nuclear forces and electroweak operators with a systematic expansion of many-body calculations~\cite{Herg20review, Hebe203NF,Ekstrom2022}. The expansion of nuclear forces is usually based on chiral effective field theory (EFT)~\cite{Epel09RMP,Mach11PR} which predicts consistent nucleon-nucleon ($NN$) interactions $V_{NN}$ and three-nucleon ($3N$) interactions $V_{3N}$. The resulting Hamiltonians are then used as starting point for a range of many-body methods that can tackle nuclei as heavy as $^{208}$Pb (see, \eg{}, Refs.~\cite{Hage14RPP,Herg16PR,Soma20SCGF,Tichai2020review,Hu2021lead}). Understanding the emergence of complex structures and quantifying the EFT and many-body uncertainties are important ongoing efforts in \ai{} calculations.

While the long-range parts of nuclear forces is governed by pion exchanges, the short-range parts are given by contact interactions with a set of low-energy couplings (LECs). The set of LECs $\{c_i \}$ is usually fitted to experimental $NN$ scattering and few-body data, including sometimes also selected medium-mass observables. The nuclear Hamiltonian in chiral EFT can then be decomposed in the form
\begin{align}
    H = T+ V_\text{long} + \sum_i c_i V_i + V_\text{rest} \,,
\end{align}
where $T$ is the intrinsic kinetic energy, $V_\text{long}$ are the long-range parts, $\vec{c} = (c_1, ...,c_N)$ denotes the vector of LECs with $V_i$ the corresponding operators, and $V_\text{rest}$ are all remaining contributions that do not depend on the LECs. Depending on the truncation in the chiral power counting, the total number of LECs can be quite sizeable, \eg{}, $23$ at next-to-next-to-next-to-leading order (\nxlo{3}), resulting in a high-dimensional parameter space.

The study of sensitivities of nuclear observables to different LECs has been of growing interest in recent years as it allows to understand which interactions drive the emergence of nuclear structure (see, \eg{},~\cite{Hu2021lead,Sun2025_multiscale}).
However, the large dimension of the LEC space makes the systematic study of the associated interactions through explicit variations of the underlying LEC values impossible, as it requires millions of many-body calculations to even partly exhaust the parameter space at any realistic chiral order.
As a remedy, emulator frameworks have been developed in the last years, as they allow approximating the full many-body solution at low computational cost. While the emulator itself induces uncertainties, its accuracy can be systematically improved by enlarging the training set that is used in its initial construction. Emulators for LEC variations have been applied to nuclear few- and many-body calculations~\cite{Ekstroem2019,Koenig2020,Furnstahl2020,Drischler2021,Zhang:2021jmi,Hu2021lead,Djaerv2022,Becker2023,Drischler:2022ipa,Sun2025_multiscale} as well as schematic low-dimensional models with phase transitions~\cite{Frame2018,Companys2021}.

One powerful emulator framework is eigenvector continuation (EC)~\cite{Frame2018,Duguet2025rmp}, which is based on a diagonalization within a manifold of training vectors. Eigenvector continuation has been explored in many applications by the nuclear physics community~\cite{Frame2018,Duguet2025rmp,Ekstroem2019,Koenig2020,Furnstahl2020,Drischler2021,Zhang:2021jmi,Hu2021lead,Companys2021,Djaerv2022,Becker2023,Drischler:2022ipa,Companys2024,Sun2025_multiscale}. In this work, we use EC-based emulators to shed light on the large mean-square radius increase from \elem{Ca}{48} to \elem{Ca}{52}~\cite{Ruiz16Calcium}, which is not reproduced in \ai{} calculations using different chiral nucleon-nucleon ($NN$) and three-nucleon ($3N$) interactions.
Because nuclear radii (or the saturation density in infinite matter) are reasonably well reproduced at the Hartree-Fock (HF) level in \ai{} calculations with softer interactions~\cite{Sun2025_multiscale,Alp:2025wjn}, we construct an EC-based HF emulator to quantify the sensitivity to the five different $3N$ couplings $c_1,c_3,c_4,c_D,c_E$ of the leading $3N$ forces in chiral EFT. For the most promising LEC solutions, we also carry out in-medium similarity renormalization group (IMSRG) calculations.

This paper is organized as follows.
Section~\ref{sec:emulator_framework} discusses the formalism of the EC-based HF emulator, while Sec.~\ref{sec:emulator_construction} provides the emulator construction and details for our calcium applications. Our physics results for \elem{Ca}{48} and \elem{Ca}{52} are presented in Sec.~\ref{sec:results}. Finally, we conclude with a summary and outlook in Sec.~\ref{sec:outlook}.

\section{Emulator framework}
\label{sec:emulator_framework}

\subsection{Eigenvector continuation}

In the EC approach, the many-body states $| \Phi(\vec{c}_i) \ra$ are computed using a given many-body method for a specified set of $\NEC$ LEC training vectors $\vec{c}_i$. Based on this training set, the emulated energies can be evaluated for arbitrary values for the LECs  by solving the generalized eigenvalue problem~\cite{Frame2018,Duguet2025rmp}
\begin{align}
    H \vec{x} = E N \vec{x} \, ,
    \label{eq:geneig}
\end{align}
with the Hamiltonian and norm kernels
\begin{subequations}
\begin{align}
    H_{ij} &= \la \Phi_i | H(\vec{c}_\circ) | \Phi_j \ra \, , \\[1mm]
    N_{ij} &= \la \Phi_i | \Phi_j \ra \, .
\end{align}
\end{subequations}
Here, $\vec{c}_\circ$ denotes the target LECs and we used the short-hand notation $|\Phi_i \ra \equiv | \Phi(\vec{c}_i) \ra$.
Hence, in the EC approach the full solution is re-expanded in terms of the basis consisting of the training vectors, and the emulated states are variationally optimized through the solution of Eq.~\eqref{eq:geneig}.
Although eigenvalues may change strongly as a function of the LECs, the eigenvectors themselves can typically be more robustly inter- or extrapolated beyond the training data~\cite{Frame2018,Koenig2020,Duguet2025rmp}.

Clearly, the power of the EC approach is directly linked to the complexity of the training vectors themselves that affect the evaluation of Hamiltonian and norm kernels needed for the solution of Eq.~\eqref{eq:geneig}.
In this work, we use training vectors obtained from a Hartree-Fock solution giving rise to a variationally optimized Slater determinant~\cite{RingSchuck80}. 
The simplicity of the training vector allows for a low-cost evaluation of operator kernels.
In general, having a good description of the target observable is necessary for constructing accurate many-body emulators. While ground-state energies are generally underbound at the HF level, nuclear radii are better reproduced and much less sensitive to correlation effects (see, \eg{},~\cite{Sun2025_multiscale}).
As this work is dedicated to trends of nuclear charge radii, HF training vectors offer a good compromise between simplicity and accuracy.

Note that in its standard formulation, the EC diagonalization generates a superposition of training vectors
\begin{align}
    | \Psi(\vec{c}_\circ) \ra = \sum_i a_i \, | \Phi_i \ra \, ,
\end{align}
where $a_i$ denotes the expansion coefficients. Since $\psitgt$ is a superposition of Slater determinants, this emulated target state is beyond the scope of a HF optimization and resembles a generator coordinate method (GCM) state that is energetically lower than its HF counterpart $|\Phi_\text{HF}(\vec{c}_\circ) \ra$.

\subsection{Transition densities}

In \ai{} calculations, nuclear interactions and states are often expanded in a harmonic oscillator (HO) basis. Let $U_i$ denote the unitary transformation from the HO to the HF basis of $|\Phi_i\ra$,
\begin{align}
    a_{i,q}^\dag=\sum_p U_{i,pq}c_p^\dag\, ,
\end{align}
where indices $p,q$ denote general single-particle states.
The matrix elements of $U_i$ are given by the set of HF coefficients.
The overlap between two different HF states can be compactly written by virtue of the simple version of the Onishi formula~\cite{Loewdin971474,Onishi1966}
\begin{align}
    \la \Phi_0 | \Phi_1\ra = \det\,(U_0 U_1^\dagger) \, ,
\end{align} 
involving their respective transformation matrices.
In our calculations, the HF solutions are constrained to maintain rotational invariance. 
Hence, the determinant is evaluated in individual sub-blocks of good total angular momentum, parity, and isospin.

The matrix elements of operator kernels are conveniently evaluated using the one-body transition density matrix defined as
\begin{align}
\rho^{[01]}_{pq} = \frac{ \la \Phi_{0} | c^{\dag}_{q} c_{p} | \Phi_{1} \ra}{\la \Phi_{0} | \Phi_{1} \ra } \, .
\end{align}
In practice the one-body transition density matrix is obtained through the use of the Thouless theorem~\cite{Thou60theorem}, 
\begin{align}
    |\Phi_1 \ra = e^Z | \Phi_0 \ra \la \Phi_0 | \Phi_1 \ra \, ,
\end{align}
where the exponential involves a skew-symmetric one-particle--one-hole excitation operator
\begin{align}
    Z = \sum_{ai} z_{ai} \, a^\dagger_{0,a} a_{0,i} = - Z^\dagger \, ,
\end{align}
that maps pairs of (non-orthogonal) Slater determinants onto each other. The matrix elements of $Z$ are obtained from the expression
\begin{align}
    z_{ai} = \sum_j (f^{hh})^{-1}_{ij} f_{ja} \, ,
\end{align}
where $f_{pq} = (U_1 U_0^\dagger)_{pq}$ and $(f^{hh})^{-1}$ denotes its inverse in the sub-block of hole states. The sums run over particle ($a$) and hole ($i$) HF states for $|\Phi_0\ra$ and over hole ($j$) HF states for $|\Phi_1\ra$. With the aid of the Thouless transformation, the transition density matrix can be written as
\begin{align}
    \rho^{[01]}_{pq} = 
    \la \Phi_{0} | c^{\dag}_{q} c_{p} e^Z | \Phi_{0} \ra \, .
\end{align}
Expanding the exponential and using Wick's theorem, we obtain a decomposition of the transition density matrix as
\begin{equation}
    \rho^{[01]}_{pq} = 
    \rho^{[00]}_{pq}
    + \sum_{ai} U_{0,ap} \, z_{ai} \, U_{0,iq} \, .
\end{equation}
Terms quadratic in $Z$ (or higher) do not contribute as they do not allow for a fully contracted operator product.
Moreover, higher-body density matrices factorize into anti-symmetric products of one-body densities as expected for product-type vacua such as Slater determinants.

\subsection{Operator kernels}
\label{sec:opkernel}

Using the above definition of transition densities, the kernel of a generic operator containing up to three-body terms involves the following contributions
\begin{subequations}
\begin{align}
\la \Phi_{0} | H | \Phi_{1} \ra_\text{1B} &= 
\sum_{pq} t_{pq}\rho^{[01]}_{qp} \, ,\\ 
\la \Phi_{0} | H | \Phi_{1} \ra_\text{2B} &= 
\frac{1}{2} \sum_{pqrs} v_{pqrs}\rho^{[01]}_{rp}\rho^{[01]}_{sq} \, , \\
\la \Phi_{0} | H | \Phi_{1} \ra_\text{3B} &= 
\frac{1}{6} \sum_{pqrstu} w_{pqrstu} \rho^{[01]}_{sp}\rho^{[01]}_{tq} \rho^{[01]}_{ur} \, ,
\end{align}
\label{eq:kernels}%
\end{subequations}
where $t_{pq}$, $v_{pqrs}$, $w_{pqrstu}$ are the matrix elements of the one-body part of the kinetic energy, two-body operators, and three-body interactions, respectively.
This expression is identical to the operator kernels present in the GCM~\cite{RingSchuck80}.
For the special case of identical bra and ket states, this yields the standard expressions of the HF expectation values.
As in this case, the EC operator kernels are Hermitian, \ie{}, $H_{ij} = H_{ji}^*$, we will refer to this as ``symmetric'' formulation of the EC framework.

In this work, however, we adopt a different strategy following Ref.~\cite{Sun2025_multiscale}.
We normal order the operator with respect to the ket reference state $| \Phi_1 \ra$, giving rise to the normal-ordered zero-body part $\la \Phi_1 | H | \Phi_1 \ra$.
Furthermore, the normal-ordered one-body part is given by
\begin{align}
    h_{pq} = 
    t_{pq} + 
    \sum_{rs} v_{prqs} \rho^{[11]}_{sr} +
    \frac{1}{2} \sum_{rstu} w_{prsptu} \rho^{[11]}_{tr}\rho^{[11]}_{us} 
    \, , 
\end{align}
which is the HF potential corresponding to the ket reference state. The final operator kernels are then evaluated according to
\begin{align}
    \la \Phi_0 | H_\text{HF} | \Phi_1 \ra =
    \la \Phi_1 | H | \Phi_1 \ra +
    \sum_{pq} h_{pq} \rho^{[01]}_{qp} \, ,
\end{align}
where the contraction is performed over the transition density.
This induces an asymmetry in the operator kernels as the normal-ordered one-body part $h_{pq}$ solely depends on the ket reference state.
Due to the broken hermiticity $H_{ij} \neq H_{ji}^*$, we refer to this approximation as ``asymmetric'' EC.

Finally, we emphasize that the simple form of the operator kernels derived here is only valid for Slater determinant training vectors that yield a decomposition of many-body density matrices as (anti-symmetric) products of the one-body density matrix.
This factorization is no longer true for more general many-body states that contain higher-order correlations and hence induce irreducible density matrices of higher particle rank.

\subsection{Emulator for general ground-state observables}

In contrast to energies other ground-state observables, such as radii, cannot be directly obtained from a solution of an eigenvalue problem like Eq.~(\ref{eq:geneig}). Instead, such observables can be emulated by computing expectation values of a given operator $O(\vec{c}_\circ)$ with respect to the emulated EC ground state:
\begin{align}
    \la O(\vec{c}_\circ) \ra
    =
    \la \Psi(\vec{c}_\circ) | O(\vec{c}_\circ) | \Psi(\vec{c}_\circ) \ra  \, .
\end{align}
As a consequence, for each additional operator a new ($\NEC \times \NEC$)-dimensional operator kernel needs to be precomputed for the final EC emulation.
The corresponding expressions are identical to those of the Hamiltonian in terms of transition densities as presented in Eqs.~\eqref{eq:kernels}.

Since the EC eigenvalue problem is solved for a state with given symmetry quantum numbers, the emulated values for the observable are bound to the same irreducible representation of the Hamiltonian's symmetry group.
As a check of our calculations, we emulated the expectation value of $\mathbf{J}^2$. As the training manifold contains only states with good total angular momentum, the emulated expectation values indeed gave $\la \mathbf{J}^2 \ra = 0$ up to numerical precision as expected from a total angular momentum eigenstate.

\section{Emulator construction}
\label{sec:emulator_construction}

\begin{figure*}[t!]
    \centering
    \includegraphics[width=\linewidth]{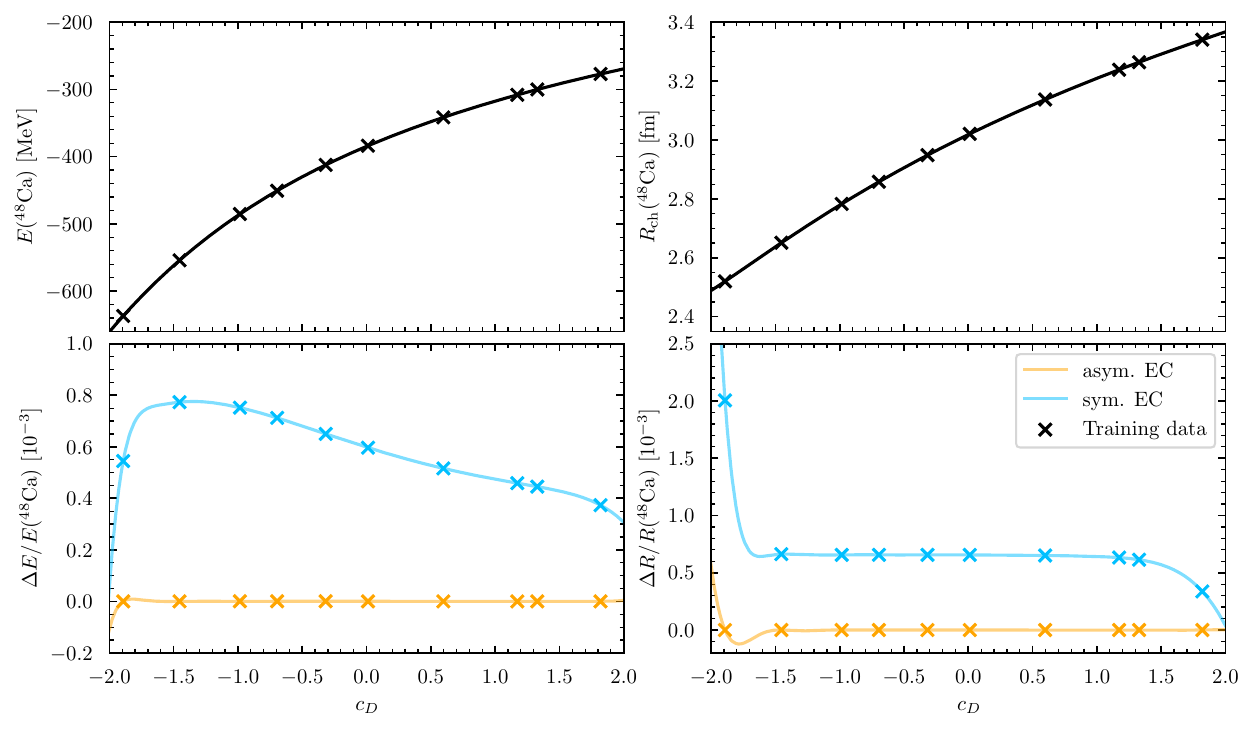}
    \caption{Emulated ground-state energies (upper left panel) and charge radii (upper right panel) together with the corresponding emulator errors (lower panels) for symmetric (blue line) and asymmetric (orange line) EC as a function of $c_D$ values. The training points are indicated by crosses. All other $3N$ LECs are fixed to the 1.8/2.0 (EM) interaction.}
    \label{fig:ECvsnonsymmetric}
\end{figure*}

\subsection{Computational details}

We consider the 1.8/2.0 (EM) $NN+3N$ interaction~\cite{Hebe11fits} for our study. The $NN$ part of this Hamiltonian is based on the N$^3$LO Entem and Machleidt (EM) interaction with cutoff $500$\,MeV~\cite{Ente03EMN3LO} which is similarity-renormalization-group (SRG) evolved to $\lambda_{\text{SRG}} = 1.8\,\mathrm{fm}^{-1}$. This is combined with a next-to-next-to-leading order (N$^2$LO) $3N$ interaction with LECs $c_1,c_3,c_4,c_D,c_E$ and non-local regulator with $\Lambda_{\text{3N}} = 2.0\,\mathrm{fm}^{-1}$. The long-range $3N$ LECs $c_i$ have been chosen consistently with the values of the $NN$ interaction, $c_1 = \bar{c}_1 = -0.81$\,GeV$^{-1}$, $c_3 = \bar{c}_3 = -3.2$\,GeV$^{-1}$, and $c_4 =\,\bar{c}_4 = 5.4$ GeV$^{-1}$, while the short-range LECs $c_D$ and $c_E$ are fit to the experimental $^{3}$H ground-state energy and $^4$He charge radius. This resulted in the values $c_D = \bar{c}_D = 1.264$ and $c_E = \bar{c}_E = -0.12$. The 1.8/2.0 (EM) interaction has proven to be very successful in describing ground-state energies of nuclei over a wide range of the nuclear chart~\cite{Simo17SatFinNuc,Morr17Tin,Stroberg2021,Arthuis:2024mnl,Bonaiti:2025bsb}. For our studies here, we will keep the $NN$ interaction fixed (we will study other $NN$ interactions in Sec.~\ref{sec:other_NN}). We vary the $3N$ LECs over a wide range of values (see Sec.~\ref{sec:training_sets}) and study the sensitivity of ground-state energies and charge radii.

The HF calculations are performed in a single-particle basis comprising 13 major oscillator shells, \ie{}, $e_\text{max}= \text{max}(2n + l) =12$, with oscillator frequency $\hbar \omega = 16 \, \MeV$ and an additional truncation on the $3N$ matrix elements $e_1 + e_2 +e_3 \leq E_\text{3max}=16$. This is typically large enough for converged calculations of $A \sim 50$ nuclei.
The HF solution is symmetry-constrained and hence is an eigenstate of total angular momentum $\mathbf{J}^2$, angular-momentum projection $\mathbf{J}_z$, parity $\mathcal{P}$ and total isospin projection $T_z$.
The $NN$ matrix elements and $3N$ matrix elements for the different LEC parts are generated in HO basis using the \nuhamil{} code~\cite{Miyagi2023EPJA_NuHamil}. The matrix elements of the nuclear Hamiltonians for the different LECs are then constructed on the fly before the many-body calculations.
The mean-square charge radius is evaluated via
\begin{align}
    \la R_\text{ch}^2 \ra = 
    \la R^2_p \ra + r^2_p + \frac{N}{Z} \, r^2_n + \la r^2_\text{so} \ra + \frac{3}{4m^2} \, ,
\end{align}
where $\la R^2_p \ra$ is the mean-square point-proton radius, $r^2_p = 0.707 \, \fm{2}$ and $r^2_n = -0.116 \, \fm{2}$ the squared proton and neutron charge radius~\cite{Workman2022PTEP_PDG2022}, $\la r^2_\text{so} \ra $ the spin-orbit correction~\cite{Ong10spinorbit,Heinz2025PRC_ImprovedCalcium}, and $3/(4m^2) = 0.033 \, \fm{2}$ the Darwin-Foldy correction (with nucleon mass $m$)~\cite{Fria97foldyShift}.

\subsection{Symmetric vs asymmetric formulation}

We start the discussion of the results by comparing the different symmetric and asymmetric formulations of the EC operator kernels presented in Sec.~\ref{sec:opkernel}.
In Fig.~\ref{fig:ECvsnonsymmetric} we show the training points and emulated results for the ground-state energy and radius of $^{48}$Ca as a function of the $3N$ LEC $c_D$ (upper panels), as well as the corresponding relative emulator errors $\Delta E/E = (E_\text{emulated}-E_\text{HF})/E_\text{HF}$ and $\Delta R/R = (R_\text{emulated}-R_\text{HF})/R_\text{HF}$ (lower panels).
As the symmetric variant is a subspace diagonalization and, therefore, yields a variational optimization of the emulated ground state in the training manifold, the emulated energy can be lower than the HF energy at the training points, since the emulated ground state is not necessarily a Slater determinant. In this case, the energy at the training points is not recovered due to the mixing with other training vectors, and the EC ground state is not a single Slater determinant.

This is different for the asymmetric EC approach, where the training state is ensured to be an eigenvector of the EC matrix at the training point and, therefore, the energy at the training points is recovered. Moreover, the emulator error of the energy is reduced by two orders of magnitude compared to the symmetric EC.
In some cases, we still observe small violations of the lower bound provided by the EC training points of the order of $\Delta E \approx 10^{-3} \, \MeV$; we attribute these to numerical effects.

As only the asymmetric formulation recovers the many-body observable at the training points, we discard the symmetric approach and from now on focus on the asymmetric formulation.
We hence drop the label ``asymmetric'' in the following and assume that EC operator kernels are always evaluated following the asymmetric EC approach as in Ref.~\cite{Sun2025_multiscale}.

\subsection{Training sets}
\label{sec:training_sets}

The LEC combinations are obtained from a $k$-dimensional latin hypercube sampling that generates a space-filling distribution of the LECs in the domain,
\begin{align}
    \mathcal{D} = [a_1, b_1] \times [a_2, b_2] \times ... \times [a_k, b_k] \, ,
\end{align}
where a single LEC $c_i$ is drawn from the interval $[a_i, b_i]$.
We choose the following large intervals for the sampling of the leading $3N$ LECs,
\begin{align}
    c_3 &\in [\bar{c}_3 - 3 \, \text{GeV}^{-1}, \bar{c}_3 + 3 \, \text{GeV}^{-1}] = [-6.2, -0.2] \: \text{GeV}^{-1} ,\notag \\
    c_4 &\in [\bar{c}_4 - 5 \, \text{GeV}^{-1}, \bar{c}_4 + 5 \, \text{GeV}^{-1}] = [0.4, 10.4] \: \text{GeV}^{-1}  ,\notag \\
    c_D &\in [-5, 10] \, ,\notag \\
    c_E &\in [-2, 2] \, ,
    \label{eq:LEC_intervals}
\end{align}
with a total number of $N_t =100$ training points to reduce the emulator error to less than $1\%$.
While the size of this training data is substantially larger than in previous applications, the regularization procedure described in Sec.~\ref{sec:reg} reduces the dimension of the remaining EC subspace to around 50-80, depending on the imposed cut of the norm matrix.

\begin{figure}[t!]
    \centering
    \includegraphics[width=\linewidth]{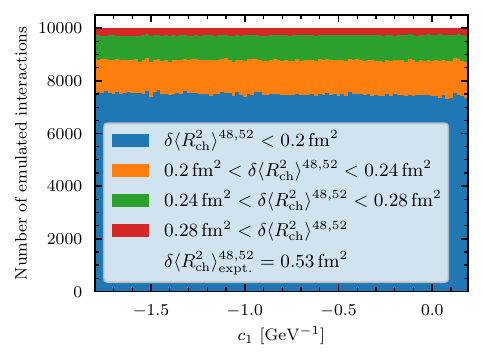}
    \caption{Distribution of different ranges for $\delta \la R_\text{ch}^2 \ra^{48,52}$ as a function of $c_1$.}
    \label{fig:c1}
\end{figure}

We keep the value of $c_1= \bar{c}_1 = -0.81$\,GeV$^{-1}$ fixed in our study because of its minor impact on charge radius trends. This is numerically demonstrated in Fig.~\ref{fig:c1}, where the differential mean-square charge radius $\delta \la R_\text{ch}^2 \ra^{48,52} = \la R_\text{ch}^2 \ra(^{52}$Ca$) - \la R_\text{ch}^2 \ra(^{48}$Ca$)$ is shown for a wide variation of $c_1$ values. We find that the total number of emulated interactions in each bin is, to very good approximation, independent of the $c_1$ value.

\subsection{Cross validation}

\begin{figure}[t!]
    \centering
    \includegraphics[width=\linewidth]{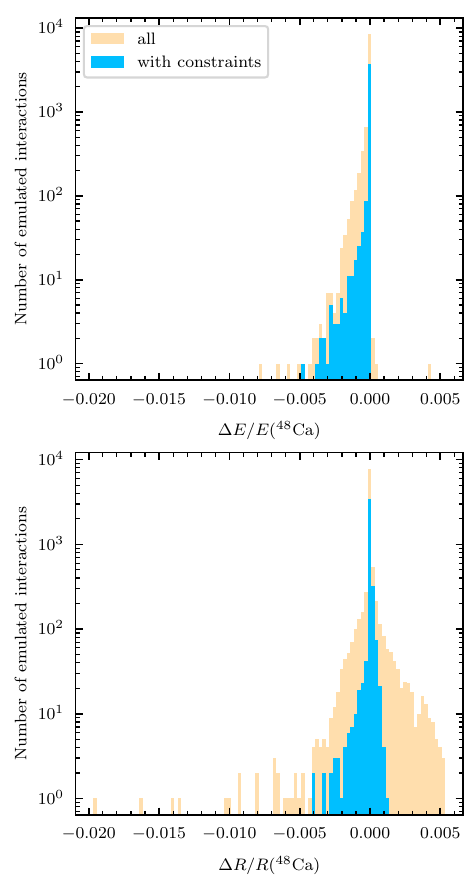}    
    \caption{Distribution of the relative error of the emulated ground-state energies compared to exact HF calculations (upper panel) and the relative error of the charge radii (lower panel). The blue part of the histograms corresponds to the interactions restricted to the energy and radius constraints from Eq.~(\ref{eq:ER_constraints}).}
    \label{fig:Crossvalidation}
\end{figure}

To ensure the accuracy of our emulator, we perform an extensive cross validation of emulated ground-state energies and charge radii, and benchmark them against the results of explicit Hartree-Fock calculations.
To this end, we set up the emulator for \elem{Ca}{48} using $N_t$ training points and generate a set of $\NEC = 10\,000$ samples for the emulator evaluation. For both we employ the same latin hypercube sampling in the LEC intervals discussed in the previous section.

In Fig.~\ref{fig:Crossvalidation}, the histograms show the number of emulated interactions as a function of the deviation from the exact HF values. The histograms are plotted using in total 100 bins over the entire error range.
The results show that the maximum deviation for our broad sample set is around $1 \%$ for energies and $2 \%$ for radii.

\begin{figure}[t!]
    \centering
    \includegraphics[width=\linewidth]{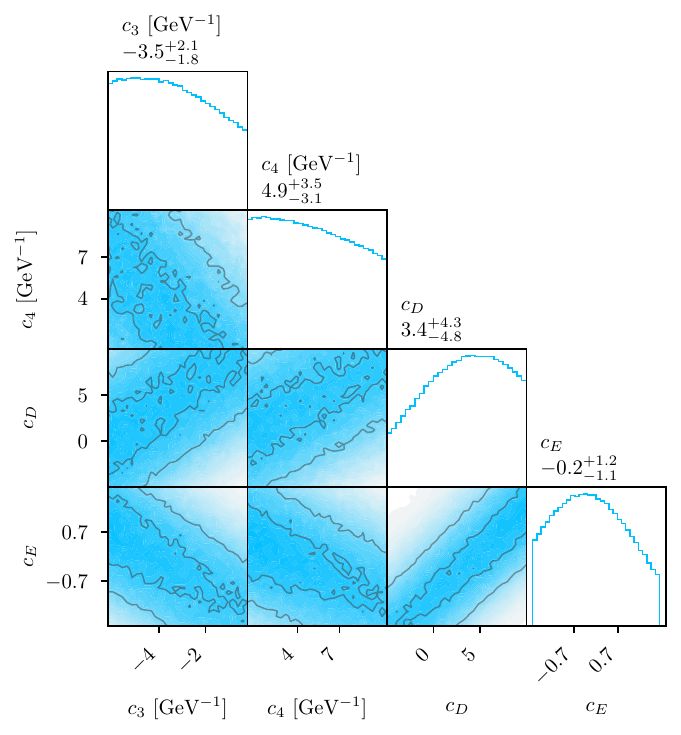}
    \caption{Distribution of the LECs for the energy and radius constraints from Eq.~(\ref{eq:ER_constraints}). Darker shades correspond to a higher number of interactions. The lines correspond to the 39.3\% and the 86.5\% levels, and the numbers for the one-dimensional distributions give the mean and 68\% confidence intervals.}
    \label{fig:Corner}
\end{figure}

We further analyze the impact of restricting the emulator points to a subset of energies and radii fulfilling the following constraints:
\begin{align}
    -466\,\MeV &< E(^{48}\text{Ca})<-100\,\MeV  , \notag \\
    2.95\,\fm{} &< R({^{48}\text{Ca}}) < 4.00 \, \fm{} .
    \label{eq:ER_constraints}
\end{align}
This corresponds to a maximal deviation of about 15\% from the experimental charge radius. For the energies the lower bound is chosen to be $50\,$MeV below the experimental value. The upper bound for the energy is chosen more conservatively since HF calculations give too-small ground-state energies compared to full \ai{} calculations. Approximately 40\% of the originally sampled set fulfill these constraints. In Fig.~\ref{fig:Corner} we show that the restricted set still covers a large part of the $3N$ LEC space.

In Fig.~\ref{fig:Crossvalidation} the corresponding interactions that fulfill these constraints are highlighted in blue, whereas all other interactions are colored in orange in the histograms. When restricting to the constrained set of emulator data, the maximum emulator error for both energies and radii is reduced significantly and does not exceed a few per mille. Given the wide range of LEC variations, the constructed HF emulator is thus remarkably precise. Note that the visualization might be a bit misleading due to the logarithmic scale in Fig.~\ref{fig:Crossvalidation}. By construction, the number of interactions that fulfill the constraints is the same for energies and radii.

\begin{figure}[t!]
    \centering    
    \includegraphics[width=\linewidth]{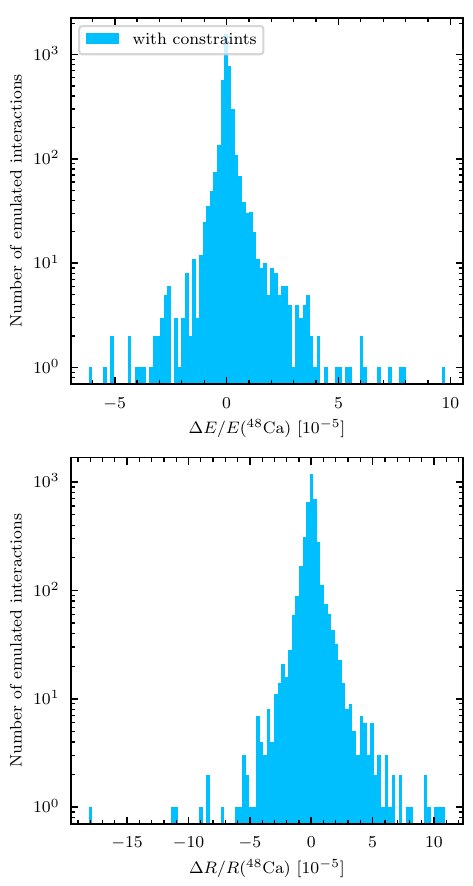}    
    \caption{Same as Fig.~\ref{fig:Crossvalidation}, but with all training vectors restricted to the energy and radius constraints from Eq.~(\ref{eq:ER_constraints}).}
    \label{fig:CrossvalidationTP}
\end{figure}

To potentially further improve the quality of the emulator for the constrained energy and radius regime, the same constraints were used on the training data in Fig.~\ref{fig:CrossvalidationTP}. In this plot only the constrained emulator results are shown. The emulator error for both energies and radii is reduced by two orders of magnitude with respect to the emulator without constraints on the training data. Outside the regime of the constrained emulated energies and radii, the accuracy worsens considerably. This is not surprising, as the training data do not cover the full LEC space anymore.
Since we are interested in Hamiltonians corresponding to realistic energies and radii, we generate our emulators based on constrained training vectors in the following, whenever possible. Note that this strategy requires calculating significantly more interactions to construct an emulator of a given size, because only around $40\%$ of the considered points satisfy the constraints.

\subsection{Regularization of the eigenvalue problem}
\label{sec:reg}

\begin{figure}[t!]
    \centering
    \includegraphics[width=\linewidth]{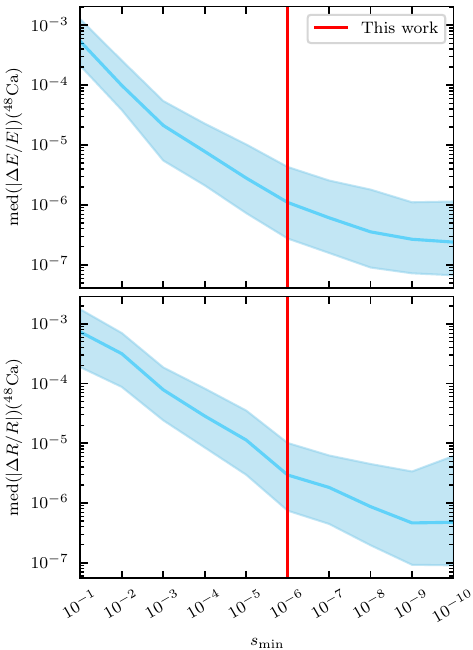}
    \caption{Median, $0.15$- and $0.85$-quantile of the cross validation as a function of the singular value cutoff $s_{\text{min}}$. The value of $s_{\text{min}}$ used in this work is marked by the red line.}
    \label{fig:Regularization}
\end{figure}

With an increasing number of training points, the training vectors tend to exhibit increasing linear dependencies, which makes the solution of the generalized eigenvalue problem Eq.~\eqref{eq:geneig} numerically unstable. This is quantified through the condition number
\begin{align}
    \kappa = \frac{\max |\lambda_i |}{\min |\lambda _i|} \, ,
\end{align}
where $\lambda_i$ are the eigenvalues of the norm matrix $N$.
The presence of small eigenvalues hence leads to a very large condition number.

\begin{figure*}
    \centering
    \includegraphics[width=\linewidth]{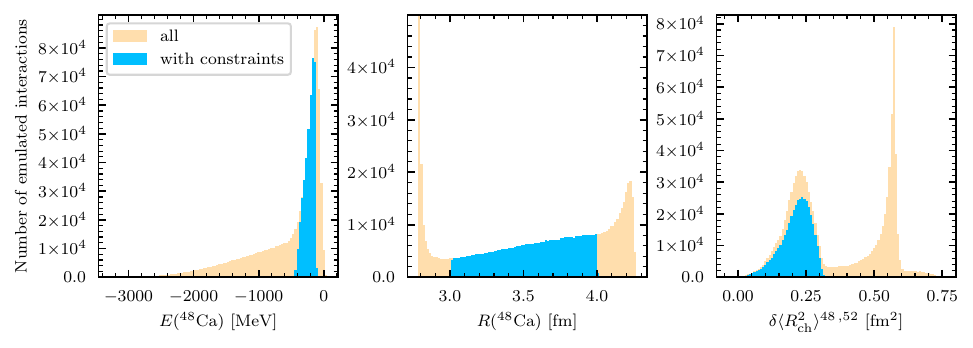}
    \caption{Distribution of emulated energies (left panel) and radii (middle panel) for $^{48}$Ca and of $\delta \la  R_\text{ch}^2 \ra^{48,52}$ (right panel). The blue distributions are restricted to the energy and radius constraints from Eq.~(\ref{eq:Ca4852constrained}).}
    \label{fig:Emulation}
\end{figure*}

As a remedy, the generalized eigenvalue problem can be regularized by projecting onto the subspace of eigenvectors with the largest norm eigenvalues.
This is practically obtained through the singular value decomposition of the norm matrix
\begin{align}
    N= L \Sigma R^T \, ,
\end{align}
where $L, R$ are orthogonal matrices and $\Sigma = \text{diag}(s_i)$ is a diagonal matrix containing the ordered set of non-negative singular values $s_i \geq 0$.
We define the subspace projection by taking the leading columns (rows) of $L$ ($R^T$) corresponding to the $k$-largest singular values with $s>s_\text{min}$. From this, we obtain a pair of nonsquare matrices $\tilde L$ and $\tilde R$ that fulfill the following orthogonality conditions in the $k$-dimensional subspace:
\begin{subequations}
\begin{align}
   \tilde L^T \tilde L = \openone_{k\times k} \, , \\
    \tilde R^T \tilde R = \openone_{k\times k} \, .
\end{align}
\end{subequations}
We then further write $\Sigma = \Sigma^{1/2} \Sigma^{1/2}$ 
and project the EC Hamiltonian according to
\begin{align}
    \tilde H = 
    \tilde\Sigma^{-1/2} \tilde L^T  H \tilde R \tilde\Sigma^{-1/2}  \, .
\end{align}
Note that $\Sigma^{-1/2}$ is well defined because $\Sigma$ is a diagonal matrix with real, positive entries.
This leads to a new eigenvalue problem in a lower-dimensional subspace
\begin{align}
    \tilde H \vec{\tilde x} = \lambda \vec{\tilde x} \, .
\end{align}
The condition number of this regularized problem is determined by the chosen $s_{\text{min}}$ value. Note that this regularization method is well known in GCM calculations that similarly suffer from linear dependencies within the set of many-body vacua generated along a set of collective coordinates~\cite{RingSchuck80,Schunk2019book}.

In Fig.~\ref{fig:Regularization} we show the sensitivity of the emulated results as a function of the chosen $s_{\text{min}}$ value. Specifically, the figure shows the median $\mathrm{med}|\Delta E/E|$ and $\mathrm{med}|\Delta R/R|$ and the $0.15$- and $0.85$-quantile of the distribution of the absolute value of the relative emulator error for the energy and charge radius of $^{48}$Ca.
While the median and standard deviation become systematically smaller down to values of $s_\text{min} \approx 10^{-6}$, the results remain stable to a good approximation below this threshold. For our results shown in the next section we use $s_{\text{min}} = 10^{-6}$.

\section{Application to calcium isotopes}
\label{sec:results}

Next, we will use our developed emulator to study 
the sensitivity of the charge radius increase from \elem{Ca}{48} to \elem{Ca}{52} to modifications of the employed nuclear interactions, focusing on the $3N$ LEC sensitivity. This is driven by the large measured charge radius increase $\delta \la R_\text{ch}^2 \ra^{48,52} = 0.530(5) \, \text{fm}^2$ by the COLLAPS collaboration~\cite{Ruiz16Calcium}. {\it Ab initio} calculations based on different chiral $NN+3N$ interactions are still not able to reproduce this large increase~\cite{Ruiz16Calcium,Heinz2025PRC_ImprovedCalcium}.

\subsection{Distribution of LEC samples}

\begin{figure}
    \centering
    \includegraphics[width=\linewidth]{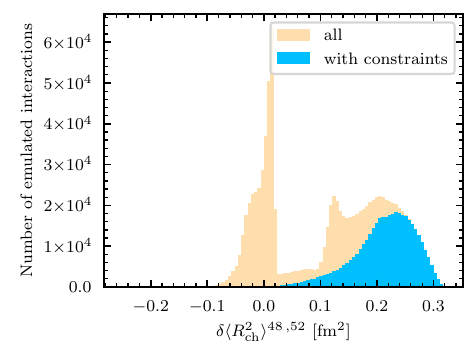}
    \caption{Distribution of the emulated $\delta\la R_\text{ch}^2\ra^{48,52}$ for the emulator with unconstrained training data, to show the difference to the emulator with constrained training data in Fig.~\ref{fig:Emulation}. The emulated blue distribution is restricted to the energy and radius constraints from Eq.~(\ref{eq:Ca4852constrained}).}
    \label{fig:Deltaunconstrained}
\end{figure}

Since we consider very large ranges for the $3N$ LECs for training and emulation, see Eq.~(\ref{eq:LEC_intervals}), not all of the considered interactions yield reasonable energies and radii. In fact, Fig.~\ref{fig:Emulation} demonstrates that a significant fraction of the considered interactions yields energies and radii very far away from the experimental regime. Because of this, we impose the following constraints on the evaluation points of the emulator:
\begin{align}
    -489 \, \MeV &< E(^{52}\mathrm{Ca})<-100 \, \MeV , \notag \\
    -466 \, \MeV &< E(^{48}\mathrm{Ca})<-100\, \MeV , \notag \\
    3.02 \, \fm{} &< R(^{52}\mathrm{Ca})< 4.09\, \fm{} , \notag  \\
    2.95 \, \fm{} &< R(^{48}\mathrm{Ca})< 4.00\, \fm{}.
    \label{eq:Ca4852constrained}
\end{align}
These ranges are the same for $^{48}$Ca as in Eq.~\eqref{eq:ER_constraints} with similar chosen boundaries for $^{52}$Ca.
As indicated for the $^{48}$Ca constraints in Fig.~\ref{fig:Corner}, the LEC distributions of the constrained interactions still cover a large part of the original LEC intervals, Eq.~(\ref{eq:LEC_intervals}).
The subset of interactions that satisfy the constraints is highlighted in blue in Fig.~\ref{fig:Emulation}. Interestingly, even when allowing for such large ranges of energies and radii, none of the $3N$ variations with the constrained energies and radii can explain the large experimental value of $\delta \la  R_\text{ch}^2 \ra^{48,52}$ (see right panel). Note that the results based on unrestricted energies and radii can have a very large emulation error. In Fig.~\ref{fig:Deltaunconstrained}, $\delta \la  R_\text{ch}^2 \ra^{48,52}$ is shown for the emulator with unconstrained training data. In this plot the large values of $\delta \la  R_\text{ch}^2 \ra^{48,52}$ for the unconstrained data disappear. The maximal value we observe in our sample set is about $\delta \la R_\text{ch}^2 \ra^{48,52} = 0.32 \,\mathrm{fm}^2$.

\begin{figure}[t!]
    \centering
    \includegraphics[width=\linewidth]{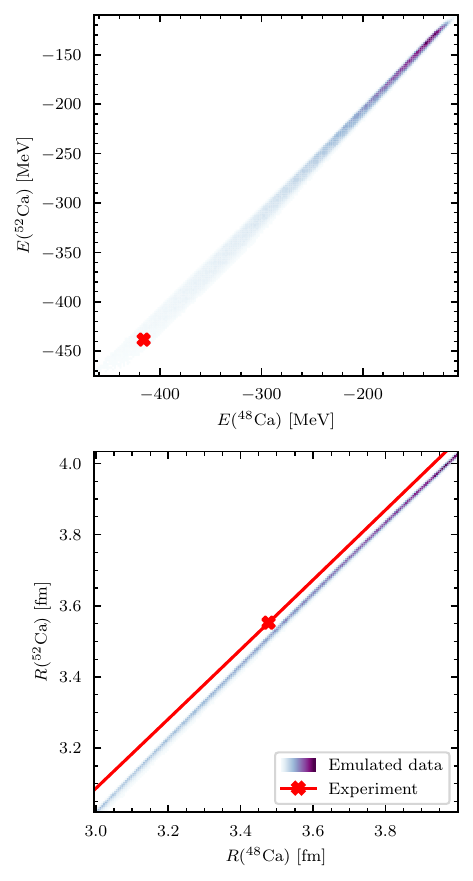}
    \caption{Distribution of ground-state energies (upper panel) and radii (lower panel) for the emulated interactions that satisfy the constraints from Eq.~\eqref{eq:Ca4852constrained}. The red crosses label experiment and the red line shows the correlation using the experimental value for $\delta\langle R_\text{ch}^2\rangle^{48,52}$.}
    \label{fig:Correlation}
\end{figure}

Finally, Fig.~\ref{fig:Correlation} shows the correlations between the ground-state energies and radii of $^{48}$Ca and $^{52}$Ca for the emulated interactions constrained by Eq.~\eqref{eq:Ca4852constrained}. For both observables, a strong correlation is observed within the variations of the $3N$ LECs. While the energy correlation band has some overlap with the experimental point (upper panel), even though only with fewer interactions, the experimental value for the radii is clearly outside of the correlation band. This shows that, at least for the considered $NN$ interaction (N$^3$LO EM 500 SRG evolved to $\lambda_{\text{SRG}}=1.8$\,fm$^{-1}$), at the HF level all $3N$ variations cannot explain the large experimental
$\delta \la R_\text{ch}^2 \ra^{48,52}$.

\subsection{Variation of the $NN$ interaction}
\label{sec:other_NN}

\begin{figure}[t!]
    \centering
    \includegraphics[width=\linewidth]{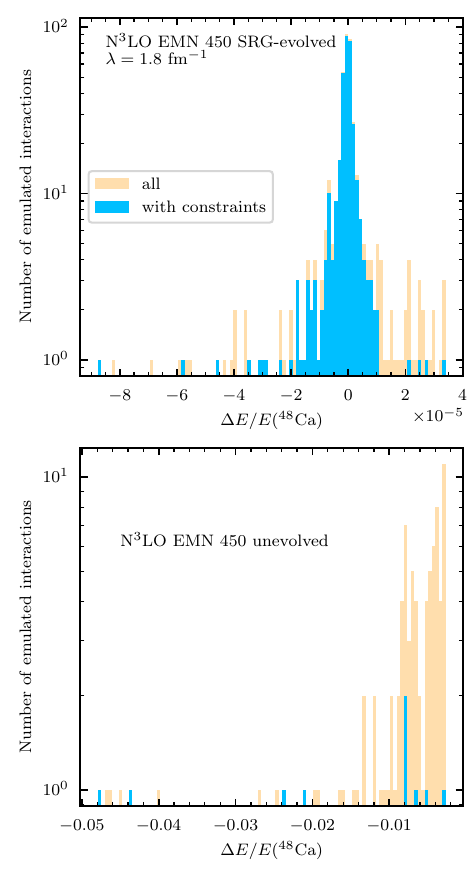}
    \caption{Distribution of the relative error of the emulated ground-state energies to exact HF calculations for different $NN$ interactions. Results are shown for the N$^3$LO EMN 450 interaction SRG-evolved to $\lambda_{\text{SRG}}=1.8$\,fm$^{-1}$ (upper panel) and unevolved (lower panel). The blue distributions are restricted to the energy and radius constraints from Eq.~\eqref{eq:Ca4852constrained}.}
    \label{fig:CrossvalidationEMN450}
\end{figure}

We therefore extend our study to other $NN$ interactions to explore the impact on the differential charge radius. In this first study, we do not vary $NN$ LECs, but instead consider the N$^3$LO Entem, Machleidt, Nosyk (EMN) interaction~\cite{Entem2017PRC_ChiralEMN} with cutoff 450\,MeV both unevolved and SRG-evolved to $\lambda_{\text{SRG}}=1.8$\,fm$^{-1}$. This has the advantage that $NN$ observables are automatically preserved. In Fig.~\ref{fig:CrossvalidationEMN450}, the cross validation of the energies is shown for $\lambda_\text{SRG} = 1.8 \, \fm{-1}$ (upper panel) and for the unevolved case (lower panel). 
We emphasize that constraining the training data for the unevolved interaction was impossible, as only about a percent of the considered data falls into the constrained energies and radii.
In the case of the unevolved interaction, the distribution restricted to the energy and radius constraints from Eq.~\eqref{eq:Ca4852constrained} is significantly wider than the distribution for the 1.8/2.0 (EM) interaction shown in Fig.~\ref{fig:Crossvalidation}. 
We suspect that the deterioration of the emulator performance is due to the appearance of unbound training states for the large $3N$ LEC intervals, which did not appear for 1.8/2.0 (EM) interaction. 
For the SRG-evolved N$^3$LO EMN 450 interaction 37\% meet the constraints, while for the unevolved case it is only about 1\%. 
Thus, not only the size of the considered LEC domain but also the actual physics on this domain impact the emulator's accuracy. 
In particular, unbound training states can lead to a significant loss in accuracy.

\begin{figure}[t!]
    \centering
    \includegraphics[width=\linewidth]{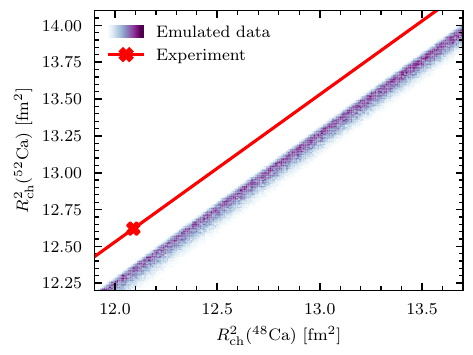}
    \caption{Distribution of the emulated mean-square chare radii for the $3N$ LEC variations with the SRG-evolved N$^3$LO EMN 450 interaction. The red line shows the correlation using the experimental value for $\delta\langle R_\text{ch}^2\rangle^{48,52}$, and the experimental point (red cross) additionally takes the experimental $^{48}$Ca charge radius into account.}
    \label{fig:Radii2EMN450}
\end{figure}

Moreover, Fig.~\ref{fig:Radii2EMN450} shows that the broad $3N$ LECs variations combined with the SRG-evolved N$^3$LO EMN 450 interaction can also not reproduce the experimental value of $\delta \la R_\text{ch}^2 \ra^{48,52}$ at the HF level. While this certainly does not represent a comprehensive sensitivity study to the employed $NN$ interactions, these results at least suggest that variations of the $NN$ interaction, under the constraints of preserving $NN$ observables, may also not yield Hamiltonians that can reproduce the large differential charge radius at the HF level.

\subsection{Impact of many-body correlations}

As a final application, we study the role of many-body correlations on the charge radii of $^{48}$Ca and $^{52}$Ca.
To this end, we employ the \ai{} in-medium similarity renormalization group (IMSRG) that accounts for the dominant particle-hole correlations in nuclear ground states~\cite{Tsuk11IMSRG,Herg16PR,Heinz2025PRC_ImprovedCalcium}. The IMSRG approach is based on a unitary transformation $U(s)$ of the nuclear Hamiltonian
\begin{align}
    H(s) = U^\dagger(s) \, H \, U(s) \, ,
    \label{eq:imsrgtrafo}
\end{align}
parametrized through a continuous flow parameter $s$.
The transformation decouples particle-hole excitations from a reference state $|\Phi \ra$, which is taken to be a HF reference state for closed-shell nuclei.
In the limit of $s \rightarrow \infty$, the IMSRG ground-state energy is given by the expectation value $E_\text{IMSRG} = \la \Phi | H(s) | \Phi \ra$.

In practice, Eq.~\eqref{eq:imsrgtrafo} is solved from the differential equation
\begin{align}
    \frac{d}{ds}H(s) = [ \eta(s), H(s)] \, ,
    \label{eq:comm}
\end{align}
with anti-Hermitian generator $\eta(s)$ that determines the unitary transformation.
Other operators, \eg{}, the charge radius, are transformed consistently with the same $U(s)$.
Solving the differential equation Eq.~\eqref{eq:comm} induces many-body operators, as the particle rank is raised through the evaluation of the commutator on the right-hand side.
These induced many-body operators must be truncated to keep the IMSRG evolution numerically tractable. 
In this work, we truncate all operators at the normal-ordered two-body level giving rise to the standard IMSRG(2) truncation, which is known to accurately reproduce ground-state properties of closed-shell nuclei~\cite{Herg16PR,Heinz2020,Heinz2025PRC_ImprovedCalcium}.
We employ the Magnus formulation of the IMSRG that allows for a more stable solution of the IMSRG flow equations through 
\begin{align}
    U(s) = \exp(\Omega(s)) \, ,
\end{align}
by solving directly for the anti-Hermitian operator $\Omega$ itself~\cite{Morr15Magnus}. This also directly yields the unitary transformation that can be used for the operator evolution.
For our calculations we use the \imsrg{} code~\cite{StrobergCode}.

\begin{figure}[t!]
    \centering
    \includegraphics[width=\linewidth]{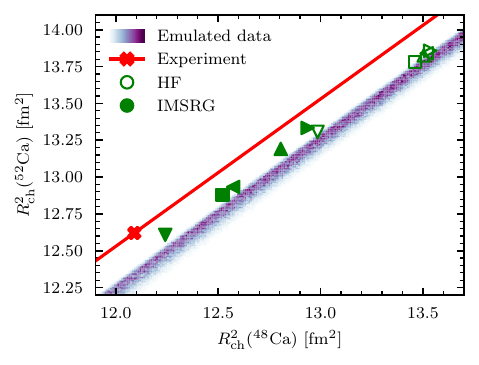}
    \caption{Distribution of the emulated mean-square charge radii for HF (open symbols) and IMSRG calculations (filled symbols). The different symbols label different selected $3N$ LECs (see text for details). Here, we have used again the $NN$ part from the 1.8/2.0 (EM) interaction.}
    \label{fig:imsrg}
\end{figure}

The large number of emulated interactions makes an exhaustive study of the impact of many-body correlations impossible. 
We therefore select a representative set of $N=20$ LEC combinations that correspond to particularly large values of the differential charge radius based on our emulated HF results and performed IMSRG(2) calculation for these cases.
In Fig.~\ref{fig:imsrg} we show the mean-square charge radius of \elem{Ca}{52} vs \elem{Ca}{48}, both at the HF and IMSRG(2) level for 5 of these interactions that were closest to the experimental values.
As in our HF survey, none of the investigated interactions was able to reproduce the large charge-radius increase. The figure also shows that it is challenging to predict the IMSRG(2) result just from the HF level, given the accuracy needed. Moreover, from our 20 interactions, the sign of the IMSRG(2) correction can be either positive or negative. Furthermore, for some of the investigated LEC combinations the IMSRG calculation did not converge, probably due to the large LEC values.
This clearly hints at the importance of including many-body correlations already in the training vectors and that too simplistic training vectors can lead to not precise enough emulators for selected observables.

\section{Summary and outlook}
\label{sec:outlook}

We have constructed an EC-based HF emulator to investigate the impact of $3N$ LEC variations on calcium isotopes, with a focus on the large charge-radius increase from \elem{Ca}{48} to \elem{Ca}{52}. 
Based on an extensive cross validation, we have demonstrated that the HF emulator has an accuracy of better than one percent.
Our results show that the charge radii of \elem{Ca}{48} and \elem{Ca}{52} are strongly sensitive to variations of the $3N$ LECs.
However, as ground-state properties are strongly correlated, the differential charge radii are less sensitive to those variations. 
As an outcome, we observed that even extensive variations of the $3N$ LECs are not sufficient to explain the large charge-radius increase towards the $N=32$ shell closure.
Exploratory variations of the underlying $NN$ interaction did also not resolve this puzzle.
This suggests that neglected many-body effects, \eg{}, associated with nuclear collectivity, play an important role.

A key challenge in our approach is the lack of many-body correlations in the training vectors. 
While many-body correlations lead to rather small changes in radii, due to our broad range of $3N$ LEC variations, it is challenging to anticipate the size (and sign) of many-body corrections to the HF radii.
This clearly motivates future extensions to use many-body data, \eg{}, IMSRG ground-state observables.
However, for the IMSRG, a wave-function-driven approach---such as EC---is not ideal, as the evaluation of operator kernels is formally challenging.
As an alternative, data-driven emulators can be used that purely rely on the observables themselves (and not on wave functions). 
Promising methods for emulating many-body data include parametric matrix models~\cite{Cook2025_pmm,Somasundaram:2024zse,Somasundaram:2024ykk,Armstrong:2025tza,Yu2025_pmm} and neural networks~\cite{Belley:2025nkn}.

\begin{acknowledgments}

We thank Gaute Hagen, Takayuki Miyagi, and Thomas Papenbrock for helpful discussions. This work was supported in part by the European Research Council (ERC) under the European Union's Horizon 2020 research and innovation programme (Grant Agreement No.~101020842) and the European Union's Horizon Europe research and innovation programme (Grant Agreement No.~101162059).

\end{acknowledgments}

\section*{Data availability}
The data that support the findings of this article are openly
available \cite{companys2026}

\bibliography{strongint}

\end{document}